\documentclass[sigconf,screen]{acmart}

\setcopyright{acmlicensed}
\acmPrice{15.00}
\acmDOI{10.1145/3368089.3409765}
\acmYear{2020}
\copyrightyear{2020}
\acmSubmissionID{fse20main-p918-p}
\acmISBN{978-1-4503-7043-1/20/11}
\acmConference[ESEC/FSE '20]{Proceedings of the 28th ACM Joint European Software Engineering Conference and Symposium on the Foundations of Software Engineering}{November 8--13, 2020}{Virtual Event, USA}
\acmBooktitle{Proceedings of the 28th ACM Joint European Software Engineering Conference and Symposium on the Foundations of Software Engineering (ESEC/FSE '20), November 8--13, 2020, Virtual Event, USA}

\usepackage{booktabs}   
\usepackage{subcaption} 

\usepackage[utf8]{inputenc}
\usepackage{graphicx}
\usepackage{url}
\usepackage{hyperref}
\usepackage{listings}
\usepackage{booktabs}
\usepackage{tabularx}
\usepackage{numprint}
\usepackage{color}
\usepackage{colortbl}
\usepackage{amsmath}
\usepackage{fontawesome}
\usepackage{microtype}
\usepackage{xspace}

\usepackage{tikz}
\usetikzlibrary{shapes}

\usepackage{enumitem}

\usepackage{arydshln}
\usepackage{booktabs}
\newcommand{\code}[1]{\texttt{#1}}

\newif\iffinal
\finaltrue

\newcommand{\tnum}[1]{{\iffinal#1\else \color{green!70!black}{#1}\fi\xspace}}
\newcounter{rqcounter}
\newcommand{\rqd}[1]{\refstepcounter{rqcounter}{\bf RQ\therqcounter\label{#1}}}
\renewcommand{\rq}[1]{{\bf RQ\ref{#1}}}

\newcounter{requirementcounter}
\newcommand{\reqDef}[1]{\refstepcounter{requirementcounter}{{\bf R\therequirementcounter\label{#1}}}}
\newcommand{\req}[1]{{\bf R\ref{#1}}}

\newcommand{\observation}[1]{
  \vspace{1mm}
  \hspace*{-\parindent}\fbox{
  \begin{minipage}{0.97\columnwidth}
    $\blacksquare$ \emph{#1}
  \end{minipage}}
}

\definecolor{gray}{gray}{0.7}
\definecolor{light-gray}{gray}{0.9}
\definecolor{KWColor}{gray}{0}
\definecolor{CommentColor}{rgb}{0.133,0.545,0.133}
\definecolor{StringColor}{rgb}{0,0.126,0.941}

\newcolumntype{g}{>{\columncolor{light-gray}}r}

\lstdefinelanguage{Pseudo}%
{morekeywords={abstract,case,catch,char,class,%
		def,else,extends,final,%
		if,import,while,for,
		do,
		match,module,new,null,object,override,package,private,protected,%
		public,return,super,this,throw,trait,try,type,val,var,with,implicit,%
		macro,sealed,%
	},%
	sensitive,%
	morecomment=[l]//,%
	morecomment=[s]{/*}{*/},%
	morestring=[b]",%
	morestring=[b]',%
	showstringspaces=false%
}[keywords,comments,strings]%

\lstdefinelanguage{Scala}%
{morekeywords={abstract,case,catch,char,class,%
    def,else,extends,final,%
    if,import,while,for,
    match,module,new,null,object,override,package,private,protected,%
    public,return,super,this,throw,trait,try,type,val,var,with,implicit,%
    macro,sealed,%
  },%
  sensitive,%
  morecomment=[l]//,%
  morecomment=[s]{/*}{*/},%
  morestring=[b]",%
  morestring=[b]',%
  showstringspaces=false%
}[keywords,comments,strings]%

\npdecimalsign{.}

\begin{document}

\title[Modular Collaborative Program Analysis in OPAL]{Modular Collaborative Program Analysis in OPAL} 



\author{Dominik Helm}
\email{helm@cs.tu-darmstadt.de}
\affiliation{
	\department{Department of Computer Science}
	\institution{Technische Universität Darmstadt}
	\country{Germany}
}

\author{Florian Kübler}
\email{kuebler@cs.tu-darmstadt.de}
\affiliation{
	\department{Department of Computer Science}
	\institution{Technische Universität Darmstadt}
	\country{Germany}
}

\author{Michael Reif}
\email{reif@cs.tu-darmstadt.de}
\affiliation{
	\department{Department of Computer Science}
	\institution{Technische Universität Darmstadt}
	\country{Germany}
}

\author{Michael Eichberg}
\email{mail@michael-eichberg.de}
\affiliation{
	\department{Department of Computer Science}
	\institution{Technische Universität Darmstadt}
	\country{Germany}
}

\author{Mira Mezini}
\email{mezini@cs.tu-darmstadt.de}
\affiliation{
	\department{Department of Computer Science}
	\institution{Technische Universität Darmstadt}
	\country{Germany}
}

\begin{abstract}
Current approaches combining multiple static analyses deriving different, independent properties focus either on modularity or performance.
Whereas declarative approaches facilitate modularity and automated, analysis-independent optimizations, imperative approaches foster manual, analysis-specific optimizations.

In this paper, we present a novel approach to static analyses that leverages the modularity of blackboard systems and combines declarative and imperative techniques.
Our approach allows exchangeability, and pluggable extension of analyses in order to improve sound(i)ness, precision, and scalability and explicitly enables the combination of otherwise incompatible analyses.
With our approach integrated in the OPAL framework, we were able to implement various dissimilar analyses, including a points-to analysis that outperforms an equivalent analysis from Doop, the state-of-the-art points-to analysis framework.
\end{abstract}



\maketitle

\section{Introduction}
\label{section:introduction}
Solving complex static analysis problems often involves solving several distinct but interdependent sub-problems.
Analyzing a method's purity, e.g., involves interdependent mutability sub-analyses of  
classes and fields~\cite{helm2018unified, porat2000automatic,huang2012reim} as well as an escape analysis~\cite{escape99choi,escape05kotzmann}. 

Traditionally, static analyses have been implemented in an imperative monolithic style, 
i.e., one super-analysis computes the results of all sub-problems.
Not only do monolithic designs become complex when mutually dependent 
problems are involved~\cite{bravenboer2009strictly}. More importantly,
individual sub-analyses cannot be developed in isolation, cannot be reused for other analyses, and cannot easily 
be added, removed, and exchanged to trade-off between precision, sound(i)ness~\cite{livshits2015defense}, and 
performance in a fine-tuned way, i.e., to enable pluggable precision/sound(i)ness/performance.

To address these requirements, it is desirable to encode solutions for sub-problems of a
complex static analysis in separate modules. 
However, while encoded in independent modules, the execution
of inter-dependent sub-analyses needs to be interleaved 
to enable exchanging intermediate results.
The latter is often necessary for optimal precision, as has been proven by the theory of reduced products in abstract interpretation~\cite{cousot1979systematic} and was more recently demonstrated for 
other kinds of analyses~\cite{bravenboer2009exception,helm2018unified,eichberg2018lattice}.

Recently, declarative approaches to static analysis using the Datalog language~\cite{whaley2004cloning,bravenboer2009strictly,madsen2016datalog} are gaining increased popularity---especially in the area of points-to analyses~\cite{whaley2004cloning,bravenboer2009strictly,smaragdakis2011pick,tan2016making}.
Such approaches nicely support the requirements stated above. 
Analyses are implemented as sets of rules that are evaluated by an underlying constraint solver.
Thus, complex analyses can be broken down into simpler, independently-developed analyses.
The underlying solver transparently resolves their dependencies and propagates 
intermediate updates according to the specified rules, thus enabling interleaved execution.
Moreover, the solver can (a) apply analysis-independent optimizations, e.g., 
by rearranging the computation order (although manual optimization is still necessary~\cite{bravenboer2009strictly,smaragdakis2010using}), and/or (b) automatically parallelize 
the execution~\cite{jordan2016souffle}. 

However, using Datalog and giving solvers full control comes with 
\emph{drawbacks in terms of both performance and generality}. 
First, it is not possible to exploit analysis-specific knowledge in managing 
the execution and dependencies of the analyses. 
Such knowledge can help boost scalability.
For example, an imperative purity analysis that determines 
whether a method is deterministic by, among others, checking the mutability of fields $f_1, ..., f_n$  could drop further checks as soon as any $f_i$ is found to be mutable.
A declarative analysis whose execution is driven by a general-purpose solver
cannot take this short-cut. Analysis-specific knowledge is also valuable to correctly 
compose incompatible optimistic and pessimistic 
analyses (as defined in~\cite{grove2001framework,lhotak2003scaling}). \
Second, the Datalog solver uses analysis-independent data structures 
and analyses cannot exploit data structures that are tailored for their specific needs.
Such optimized data structures, like tries, can be crucial for 
achieving performance.
Finally, the fully declarative approach fosters a one-size-fits-all style, limiting generality.
For instance, by relying on relations, Datalog-based approaches support only set-based lattices, 
while many common analyses require other kinds of lattices.
Constant propagation, e.g., is usually implemented via singleton-value-based lattices, 
making it infeasible to implement it using Datalog~\cite{madsen2016datalog, szabo2018incrementalizing}.

In this paper, we address these issues of declarative approaches, without comprimizing on their benefts.  
Specifically, we propose a novel generic approach together with a proof-of-concept implementation in the OPAL framework~\cite{eichberg2014software} for 
lattice-based fixed-point computations with support for lattices of any kind including singleton-value-based, interval, and set lattices.
Like fully declarative approaches, it features modular analyses encoded as independently compilable, exchangeable, and extensible units.
However, it does not rely on a general-purpose declarative framework and constraint solver.
It offers a specialized approach mixing imperative and declarative styles.
The developer of an OPAL analysis implements its core functionality imperatively, but declaratively
specifies its dependencies, e.g., the lattice that the analysis computes and lattices 
it depends on to do so, as well as several constraints regarding its execution. 
Dependencies and constraints are automatically handled by our custom 
solver during analysis execution.

Our architecture is reminiscent of blackboard systems~\cite{corkill1991blackboard}:
Dependent analyses implemented in decoupled modules coordinate their executions implicitly by writing into and reading from a central data structure (the~"blackboard").
Whenever new (intermediate) results are written to the blackboard, the solver automatically (and concurrently) schedules the execution of dependent analyses to satisfy the declaratively specified dependencies and constraints.

Like declarative approaches, we decouple mutually dependent analyses, 
enabling their isolated development and interleaved parallel execution out-of-the-box. 
At the same time, we improve over declarative approaches in two regards.
First, beyond automatic and transparent optimizations and parallelization, 
by featuring an imperative programming style within each analysis module, 
our approach enables analysis-specific optimizations and data structures.
The possibility to specify fine-grained (analysis-specific) constraints 
enables further optimizations, e.g., suppressing interleaved execution of some 
analyses to avoid unnecessary intermediate computations.
Second, with a custom solver that is agnostic of the lattices used by analyses, our approach is generic and supports arbitrary kinds of analyses.
Using it, one can naturally express dataflow and constraint-based analyses based on arbitrary lattices.
Moreover, declarative declarations enable OPAL to 
consider analysis-specific constraints in managing dependencies.
To the best of our knowledge, this is the first approach to correctly compose lazily computed incompatible optimistic and pessimistic analyses.\\

\noindent
To recap, this paper contributes:
\begin{itemize}
\item A list requirements on frameworks for collaborative static analysis that is distilled from \tnum{three} case studies (Section~\ref{section:case_study}).
\item A novel approach, that satisfies all these requirements (Section~\ref{section:approach}). It advances the state-of-the-art in implementing modular inter-dependent analyses.
\item A thorough evaluation of the approach that supports our claims on generality, showcases its modularity features, points out performance improvements over Doop~\cite{bravenboer2009strictly}, the state-of-the-art declarative framework, and provides promising results for parallelization (Section~\ref{section:evaluation}).
\end{itemize}
\section{Background and Terminology}
\label{section:background}

In this section, we shortly introduce blackboard systems and present terminology used throughout the paper.\\

\emph{Blackboard Systems}~\cite{corkill1991blackboard} use a central data structure - \emph{the blackboard} - to coordinate the collaborative work of otherwise decoupled \emph{knowledge sources}. 
The latter contribute (partial) information to the blackboard towards collaboratively reaching an overall goal.
The blackboard notifies knowledge sources about availability of new information they might require 
through a control mechanism that decides which knowledge sources should be executed in what order.
The information can then 
be queried by the knowledge sources, which execute and produce further information.
Each execution of a knowledge source is called an \emph{activation}.\\

\emph{Entity:} The term is used to characterize anything one can compute some information for.
Entities can be concrete code elements, e.g., classes, methods, or allocation sites,
or abstract concepts such as all subtypes of a class.
The set of entities is not necessarily defined a priori and can be created on-the-fly while analyses execute. 

\emph{Property Kind:} The term characterizes a specific kind of information that can be computed for an entity,
e.g., mutability of classes, purity of methods, or callees of a specific method.
Each property kind represents one lattice of possible values. 

\emph{Property:} The term characterizes a specific value in the lattice of some property kind that is attached to some entity, 
e.g., a class can be mutable or immutable, a method can be pure or impure, a specific method may invoke a specific set of methods.
Per entity at most one property of a specific kind can be computed.

\emph{Analysis:} The term characterizes an algorithm that 
given an entity computes its property of a certain kind.
We say that \emph{an analysis computes a property kind} as shorthand for "an analysis computes 
properties of that property kind for a given kind of entity".
Analyses are knowledge sources in the sense of the blackboard architecture;
the properties they compute constitute the information that they contribute to and/or query from the blackboard.

\section{Case Studies}
\label{section:case_study}

We 
discuss case studies involving several interrelated sub-analyses
to distill a list of requirements on static analysis frameworks.
During the discussion, we \emph{emphasize} concepts whenever they occur.
The case studies represent very dissimilar kinds of analyses.
In particular, they require different kinds of lattices, including singleton-value lattices (e.g.\ in \ref{sec:purity}) 
and set-based lattices (e.g.\ in \ref{sec:call_graph}).
This motivates the first requirement:
Static analyses frameworks must support varied domain lattices (\reqDef{lattices}).

\subsection{Three-Address Code}
\label{sec:tac}
The first case study is an analysis to produce a three-address code representation (TAC) of JVM bytecode, presented in more detail in previous work~\cite{reif2020tacai}.
In its basic version, TAC uses def/use, type, and value information (including constant propagation) provided by an abstract-interpretation-based analysis (AI).
To increase precision, AI may be enhanced with analyses that refine type and the value information for method return values and fields.
However, such additional analyses may negatively affect the runtime.
Hence, systematic investigation of the precision/performance trade-off is needed to decide whether to use such additional analyses on a case-by-case basis.
To this end, a separation into modules that can be enabled/disabled is beneficial.
In general, we derive the following requirements regarding support for modular pluggable analyses.

For systematically studying precision/soundness/performance trade-offs, static analysis frameworks should support 
en/disabling inter-dependent analyses (\reqDef{pluggability}). To maximize pluggability, analyses should
be defined in decoupled modules, and yet be able to collaboratively compute properties 
(\emph{collaborative analyses}).
As individual analyses can be disabled, it should be possible to specify soundly 
over-approximated \emph{fallback values}\footnote{To minimize the effect of fallback values on precision, it makes sense to compute the fallback by using locally available information, e.g., using declared type information, 
instead of always returning the same over-approximated value.} for the properties they compute, 
in lack of actual results (\reqDef{fallbacks}).

Moreover, an approach for modular collaborative analyses should support their \emph{interleaved execution}
without them knowing about each other's existence (\reqDef{interleaving}).
Two analyses are executed interleaved, if they can interchange \emph{intermediate results}.
This is important for optimal precision \cite{cousot1979systematic}: 
knowledge gained during the execution of analysis $A_1$ may be used by 
the execution of another analysis $A_2$ on-the-fly to refine its result and, in turn, 
this may enable further refinement for $A_1$. 
The precision of field- and return-value refinement analyses 
profits from interleaved executions,
as they depend on each other cyclically. 
If a method \code{m} returns the value of a field \code{f}, then \code{m}'s return value depends on \code{f}'s value.
If the value returned by \code{m} is written into \code{f}, 
then \code{f}'s value also depends on \code{m}'s return value. 

However, interleaved execution 
must in specific cases 
be suppressed to ensure correctness. 
This is the case for
the composition of 
\emph{pessimistic} and \emph{optimistic} analyses.
Pessimistic analyses start with a sound but potentially imprecise assumption and eventually refine it.
Optimistic analyses start with an unsound but (over)precise assumption and progress 
by reducing (over)precision towards a sound result. 
Field- and return-value refinement analyses are pessimistic---the declared return type of method $m$, say \texttt{List},
is a sound but eventually imprecise initial value for the return-value 
analysis; during the execution, the analysis may find out that $m$ 
actually returns the more precise result, say \texttt{ArrayList}.
AI is an optimistic analysis---it starts with the unsound assumption that all code is dead and refines 
it by adding statements found to be alive towards a sound, 
but potentially less precise result. 
Optimistic and pessimistic analyses are \emph{incompatible} for interleaved execution,
because they refine the respective lattices in opposite directions. 
As a result, exchanging intermediate results may cause inconsistencies,  
thereby violating monotonicity. 
Thus, the analysis framework must enforce that 
only \emph{final results} of pessimistic analyses
are passed to dependent optimistic analyses (and vice-versa), avoiding
interleaving and \emph{suppressing} non-final updates (\reqDef{optivspessi}). 

For illustration, consider the example of some piece of code, say $c$, 
that contains a call to a method $m_1$ that is mutually recursive with a method $m_2$, but is conditioned on a field $f$ 
being an instance of \texttt{LinkedList}. 
To analyze $c$, 
we combine a field-value analysis $FA$, an $AI$ analysis, and a call graph construction algorithm, $CG$.  
Assume that $FA$, which is a pessimistic analysis, initially reports the type of the field $f$ to be \texttt{List}. 
Given this information,  AI would optimistically consider $c$ to be live and $CG$, 
hence, will consider both $m_1$ and $m_2$ to be reachable.
Because of the mutual recursion (and also because of the monotonicity requirement), 
this result cannot be changed later, if $FA$ finds out that $f$ can only contain \texttt{ArrayList}s. 
If, however, the latter information was present when AI analyzed the code,
$c$ would have been marked as dead, and $CG$ would have marked
$m_1$ and $m_2$ as unreachable.
Thus, the results of this combination of analyses is non-deterministic and possibly incorrect (imprecise, if $m_1$ and $m_2$ are falsely reported to be reachable).

\subsection{Modular Call Graph Construction}
\label{sec:call_graph}

Inter-procedural analyses presume a call graph (CG): Given method \code{m}, CG provides 
information about (a) methods that may be invoked at a call site in \code{m} (callees) and (b) 
call sites from which \code{m} may be invoked (callers).
We use the CG to motivate the need for supporting further 
kinds of execution interleaving 
(beyond \req{interleaving}) 
as well as further requirements. 
The previous case study illustrated the need for interleaved execution of inter-dependent analyses 
that calculate different properties and operate on different entities 
(composition of analyses for refining field and return values with TAC). 
The CG use case illustrates two further kinds of interleaved execution. 

First, we need interleaved execution of multiple instances of the same analysis operating on different code entities to collaboratively compute a single property, whereby each instance contributes partial results (\reqDef{partial}).
For example, different executions of a CG analysis for different callers of a method $m$ 
need to contribute their \emph{partial results} to collaboratively derive all of $m$'s callers
(computing callers of a method is inherently non-local).

Second, 
we also need 
to support interleaving of independent analyses that collaboratively 
compute a single property 
(\reqDef{partialcollab}).
Consider, e.g., the computation of the callees of \code{m}.
A CG analysis can in principle consider $m$ in isolation. 
A monolithic analysis for callees is nonetheless not 
suitable. 
It makes sense to distinguish between one sub-analysis 
that handles standard invocation instructions
(e.g., CHA~\cite{dean1995optimization}, 
RTA~\cite{bacon1996fast}, points-to-based~\cite{bravenboer2009strictly} analysis)  
and sub-analyses dedicated to non-standard ways of method invocation through
specific language features, e.g., reflection, native methods, or functionality 
related to threads, serialization, etc. Non-standard invocation 
requires specific handling (e.g., one may deliberately not want to 
perform reflection resolution, or may want 
to perform it based on dynamic execution traces).
By offering such specialized analyses as decoupled modules, they become highly reusable
and can be combined with different call-graph analysis for standard invocation instructions. 
This makes the call graph construction highly configurable
for fine-tuning its performance and sound(i)ness.
Hence, not only a method's callers but also its callees need to be computed collaboratively.
This time, different analyses targeting different language features, rather than different executions of the 
same CG analysis, contribute to the same property.

Handling special language features may even rely on integrating
results of external tools or precomputed values (\reqDef{external}).
For instance, one may choose to integrate the results of
TamiFlex~\cite{bodden2011taming} for reflective calls,
or external tools for analyzing native methods.

The CG case study also motivates support
for specifying precise \emph{default values} (\reqDef{defaults}) (in addition to 
sound fallback values).
Consider the case of an unreachable method $m$.
The CG analysis will never compute callees or caller information for $m$.
However, this lack of results is an inherent property of the entity and not the result of a missing/disabled analysis.
A sound fallback value to compensate the deactivation of the CG module for $m$
may have to include all methods and hence be too imprecise.
Instead, analyses depending on the CG should get the information that 
$m$ is unreachable---the precise default value.
The analysis developer knows such information and should be enabled to tell the framework. 

Another requirement is motivated by the CG.
The CG construction unfolds along the transitive closure of methods reachable from some entry points.
Hence, it does not make sense to execute the decoupled modules collaboratively constructing the CG---each handling a particular language feature---globally on all methods of a program.
Instead, they should be \emph{triggered} only when the overall analysis progress discovers a newly reachable method.
Hence, the framework must support triggering analyses once the first (intermediate) result for a property is recorded (\reqDef{triggered}).

Our previous work~\cite{reif2019judge} provides empirical evidence that 
encoding an RTA sub-analysis and sub-analyses for language-specific features as collaborative interleaved modules, 
results in more sound call graphs and better performance compared to call graph analyses of the Soot~\cite{vallee2010soot}, WALA~\cite{walaIBM}, and Doop~\cite{bravenboer2009strictly} frameworks.

\subsection{Mutability, Escape, and Purity Analysis}
\label{sec:purity}

The analyses in this subsection illustrate the need for further kinds 
of activation modes in addition to triggered analyses, illustrated in the previous subsection:
(a) \emph{eager analyses},
which refers to computing an analysis for all entities 
in the analyzed program, and (b) \emph{lazy analyses},
i.e., executing an analysis $A_1$ only for the entities for which the property that $A_1$ computes is queried by 
some (potentially the same) analysis $A_2$. A further requirement shown by analyses 
in this subsection is that the framework should allow analyses to enforce an execution order 
that overrides the one determined by the solver.
 
The use case involves analyses for method purity, class and field mutability~\cite{porat2000automatic,huang2012reim}, 
and escape information~\cite{escape99choi,escape05kotzmann}.
The latter includes aggregated information on field locality and return-value freshness (cf.~\cite{helm2018unified}).
The analyses in this case study interact tightly and compute properties that 
may be relevant for both end users (e.g., method purity) and further 
analyses (e.g., escape information).
Complex dependencies exists between all these analyses.
To fine-tune the precision/performance trade-off, several analyses for 
these property kinds with different precision can be exchanged as needed; 
all are \emph{optimistic} and use TAC and/or the CG information.

Since the results of analyses in this case study may be of interest to 
the end user, it is useful to compute them for all possible entities eagerly
(\reqDef{eager}), 
e.g., computing the mutability of all fields in the program.
However, when the field mutability is only used to support, e.g., 
the purity analysis, it may be beneficial for performance reasons to compute it
lazily (\reqDef{lazy}), i.e., only for the fields for which mutability
is queried by the purity analysis.
This illustrates that we need both eager and lazy execution modes. 
Eager and lazy versions of the same analysis should typically share the 
code and only be registered with the framework in different ways.
The class mutability analysis also illustrates the need to configure the 
framework with analysis-specific execution orders (\reqDef{incremental}):
For performance reasons, it makes sense to analyze classes in a program
in a top-down order starting with parent classes before their children.

Our previous work (\cite{helm2018unified}) provides empirical evidence for the requirements stated in this section.
An implementation of the purity sub-analysis of this case study (and through transitive use, the mutability and escape sub-analyses)
as collaborative analyses with interleaved execution showed higher precision, more fine-granular results and similar performance characteristics compared to the then state-of-the-art purity inference tool ReIm~\cite{huang2012reim}.

\begin{table}
	\small
	\centering
	\caption{Summary of Requirements}
	\begin{tabularx}{\columnwidth}{p{0.3cm}l}
		\toprule
		\multicolumn{2}{l}{\emph{Lattices and values}}\\
		\req{lattices} & Support for different kinds of lattices (\ref{sec:tac}, \ref{sec:call_graph}, \ref{sec:purity})\\
		\req{fallbacks} & Fallbacks of properties when no analysis is scheduled (\ref{sec:tac}, \ref{sec:purity})\\
		\req{defaults} & Default values for entities not reached by an analysis (\ref{sec:call_graph})\\
		\midrule
		\multicolumn{2}{l}{\emph{Composability}}\\
		\req{pluggability} & Support for enabling/disabling individual analyses (\ref{sec:tac}, \ref{sec:call_graph}, \ref{sec:purity})\\
		\req{interleaving} & Interleaved execution with circular dependencies (\ref{sec:tac}, \ref{sec:call_graph}, \ref{sec:purity})\\
		\req{optivspessi} & Combination of optimistic and pessimistic analyses (\ref{sec:tac})\\
		\req{partial} & Different activations contributing to a single property (\ref{sec:call_graph})\\
		\req{partialcollab} & Independent analyses contributing to a single property (\ref{sec:call_graph})\\
		\midrule
		\multicolumn{2}{l}{\emph{Initiation of property computations}}\\
		\req{external} & Precomputed property values (\ref{sec:call_graph}, \ref{sec:purity})\\
		\req{triggered} & Start computation once an analysis reaches an entity (\ref{sec:call_graph})\\
		\req{eager} & Start computation eagerly for a predefined set of entities (\ref{sec:purity})\\
		\req{lazy} & Start computation lazily for entities requested (\ref{sec:tac}, \ref{sec:purity})\\
		\req{incremental} & Start computation as guided by an analysis (\ref{sec:purity})\\
		\bottomrule
	\end{tabularx}
	\label{tbl:req}
\end{table}

\subsection{Interim Summary}
\label{sec:req}
Table~\ref{tbl:req} summarizes the requirements along the case studies motivating them.
Existing frameworks do not satisfy all of them.
Imperative frameworks lack support for modularity, especially \req{optivspessi}, \req{partial}, and \req{partialcollab}.
Declarative approaches, e.g., Doop~\cite{bravenboer2009strictly}, have other limitations:
Being bound to relations for modeling properties, they
can not express the range of different analyses represented by our case studies (\req{lattices}).
They also fail to support sound interactions between incompatible analyses (\req{optivspessi}). 
By giving the solver full control, they do not support different analysis-specific activation modes (\req{triggered}-\req{incremental}).

\section{Approach}
\label{section:approach}

OPAL is the first static analysis framework to build upon the concept of blackboard systems:
Static analysis modules correspond to knowledge sources; 
the store that manages the computed properties corresponds to the blackboard.
OPAL combines imperative and declarative programming styles for analyses.
The developer of an analysis \code{A}: 
(a) implements the lattice representation of the property values computed by \code{A}
(\ref{subsec:properties}), 
(b) implements two \emph{imperative} functions 
- so-called \emph{initial analysis function} (IAF) respectively \emph{continuation function} (CF) (\ref{subsec:analysis_structure}),
(c) declares the property kinds computed by \code{A} and properties \code{A} depends on (\ref{subsec:lifecycle}), 
and 
(d) defines how \code{A}'s results are reported to the blackboard (\ref{subsec:results}).
Guided by the declared dependencies, the blackboard and its fixed-point solver coordinate 
the execution of the analyses, thereby (e) ensuring all execution constraints (\ref{subsec:execution_constraints}), (f) 
performing fixed-point computations, whenever circular dependencies are involved (\ref{subsec:fixed-point}), and (g)
automatically scheduling and parallelizing the execution of analyses (\ref{subsec:approach_parallelization}). 

\subsection{Representing Properties}
\label{subsec:properties}

Values of a property kind constitute a lattice structure. OPAL supports
singleton value-based, interval, or set-based lattices are possible (\req{lattices}).
A lattice's bottom value models the best possible value (e.g., pure for method purity); 
its top value the sound over-approximation (e.g., impure).
Lattices must satisfy the ascending (descending) chain condition to ensure termination of optimistic (pessimistic) analyses.
When defining a property kind, developers can choose the most suitable data structures for efficiency.

Developers can also specify \emph{fallback} and \emph{default} values.
The blackboard will return the \emph{fallback value} for some requested property, \code{p} of kind \code{k}, if no analysis is available for \code{k} (\req{fallbacks}).
As it is a sound over-approximation, the lattice's top value is a good choice - 
however, the fallback value can also be provided by a "proxy" analysis function 
that does not query the blackboard, avoiding cyclic dependencies.
The blackboard will return a \emph{default value} for \code{p}, if an analysis is available, but 
did not produce any result for some
entity (\req{defaults}). For instance, 
call graph analyses only examine methods reachable from entry points - for any non-reachable method, \code{m},
a default value can be used to state that \code{m} is dead and has no (relevant) callees.
A sound fallback value would include all possible methods as callees of \code{m}; thus, 
in this case, the default value provides more information than a fallback value.
If no default value is declared, the fallback value is returned.

\begin{figure}
\begin{lstlisting}[numbers=left, numberstyle=\scriptsize\color{gray}\ttfamily,xleftmargin=2em, caption={Class Mutability Lattice}, label={lst:mutability_lattice}]
sealed trait ClassMutability extends PropertyKind {
	def fallback(Type theClass) = MutableClass
}
case object ImmutableClass extends ClassMutability
case object MutableClass extends ClassMutability
\end{lstlisting}
\end{figure}

Developers implement property kinds by specifying an interface, which can be used to access and manipulate the property values.
When the \code{PropertyKind} trait is extended, the framework assigns an identifier, which can be used to query the blackboard for 
properties of that kind.
Listing~\ref{lst:mutability_lattice} shows exemplary Scala code of a simple class mutability property kind.
Lines \tnum{1} to \tnum{3} define the base trait for the property kind and give a sound fallback value in line \tnum{2}.
The two possible property values are defined in lines \tnum{4} and \tnum{5}. 

\subsection{Analysis Structure}
\label{subsec:analysis_structure}

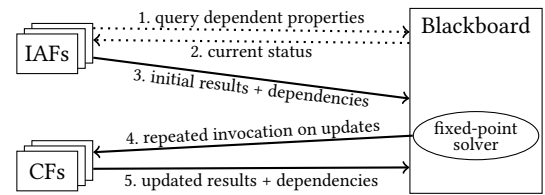
\begin{figure}
	\centering
	
	\tikzset{
		iaf/.style={anchor=north,rectangle,fill=white,draw=black,minimum width=23pt,minimum height=15pt},
		query/.style={->,dotted,thick}
	}
	
	\begin{tikzpicture}[]
	
	\node[iaf] (iaf1) at (0.20, 0) {};
	\node[iaf] (iaf2) at (0.10, -0.07) {};
	\node[iaf] (iaf3) at (0, -0.14) {IAFs};
	
	\node[anchor=north,rectangle,draw=black,minimum width=50pt,minimum height=70pt,label={[anchor=north]Blackboard},align=center] (blackboard) at (5.7, 0.2) {};
	
	\node[ellipse,draw=black,inner sep=1pt,align=center] (fps) at (5.7,-1.5) {\footnotesize fixed-point\\[-5pt]\footnotesize solver};
	
	\draw[query] (iaf1.26) to node[auto,label={[label distance=-9pt]\footnotesize 1. query dependent properties}]{} (blackboard.134);
	
	\draw[query] (blackboard.139) to node[below,label={[label distance=-9.5pt]\footnotesize 2. current status}]{} (iaf1.06);
	
	\draw[->,thick] (iaf1.335) to node[below,label={[label distance=-16pt]\rotatebox{-6.5}{\footnotesize 3. initial results + dependencies}}]{} (blackboard.178);
	
	\node[iaf] (cf1) at (0.20, -1.56) {};
	\node[iaf] (cf2) at (0.10, -1.63) {};
	\node[iaf] (cf3) at (0, -1.7) {CFs};
	
	\draw[->,thick] (fps.180) to node[above,label={[label distance=-12pt]\rotatebox{3}{\footnotesize 4. repeated invocation on updates}}](lab){} (cf1.15);
	
	\draw[->,thick] (cf1.345) to node[auto,label={[label distance=-18.5pt]\footnotesize 5. updated results + dependencies}]{} (blackboard.225);
	\end{tikzpicture}
	
	\caption{Overview}
	\label{fig:overview}
\end{figure}

An overview of OPAL's analysis structure is shown in Figure~\ref{fig:overview}.
As mentioned, the analyses are structured in two parts: 
An \emph{initial analysis function (IAF)} and one or more \emph{continuation functions (CFs)}.
These functions can be implemented in any way, as long as they provide their results as defined by the property kind.

For each entity \code{e} to be analyzed by \code{A}, \code{A}'s IAF is executed.
The IAF collects information directly from \code{e}'s bytecode in order to compute its result.
If it needs additional information pertaining to some other entity \code{e} 
or from another analysis that computes a property kind \code{k}, the IAF queries the blackboard 
for these dependencies, using the identifiers of \code{e} and \code{k} to find the relevant information (arrow 1. in Figure~\ref{fig:overview}).
The blackboard will return the currently available value (2.).
This value may, however, not be available, or not final, either
because the respective analysis was not yet executed or 
because it has dependencies that yet need to be satisfied.
Once the IAF completes analyzing the entity, it returns to the blackboard 
(a) a result computed based on the currently available information
and (b) any remaining dependencies, along with a continuation function (CF) (3.).
Similar to the solver of \emph{declarative} frameworks, the blackboard
resolves dependencies and automatically invokes the CFs whenever updates to these dependencies
become available (4.).
On completion, CFs also return their updated results to blackboard (5.), potentially triggering the execution of other CFs.
While the IAF is written imperatively (dotted queries in Figure~\ref{fig:overview}), the subsequent execution is performed similar to declarative frameworks (straight lines)
by having results declare their dependencies and the solver being responsible to satisfy them.
Executions of the IAFs and CFs are called \emph{analysis activations}.
To ensure determinism, OPAL executes the activations for a single property 
sequentially, while IAFs and CFs for other properties can execute concurrently. 

As analyses get notified about dependency updates through the invocation of the CF, 
it is not necessary that dependencies are computed before or when they are queried.
Instead, they can be computed asynchronously and lazily, i.e., on-demand (\req{lazy}).
This also allows OPAL to handle cyclic dependencies (\req{interleaving}).

Apart from adhering to this basic structure, developers may use any suitable strategy to implement an analysis \code{A}.
\code{A} may, e.g., focus on specific statements instead of traversing the entire code of a method 
(OPAL provides pre-analyses to query specific parts of the code, e.g., all statements that access a specific field).
Also, analyses can internally use any data structure suitable to achieve good performance.
For illustration, Listing~\ref{lst:example_analysis} shows an excerpt from a simple class mutability analysis' initial analysis function.
The IAF is given the entity to analyze (Line \tnum{1}).
Lines \tnum{3} to \tnum{7} show how to retrieve and handle properties required to compute the IAF's result:
The required property (the mutability of an instance field of the analyzed class) is queried from the blackboard (line \tnum{3}) and based on the returned value, the IAF may compute its result (as in line \tnum{4}) or keep the dependency in a list of dependees (line \tnum{6}) to return it alongside an intermediate result later (line \tnum{9}).
Line \tnum{9} also specifies the continuation function to be invoked when any of the properties in \texttt{dependees} is updated.
We do not show the code for that CF here, as its implementation is very similar to lines \tnum{4} to \tnum{9}, i.e., based on the updated value, the (intermediate) result of the CF is determined.

There are two semantic constraints that the implementations of the analyses must satisfy, though.
First, they must ensure \emph{monotonicity of result updates} according to the used lattice.
Analyses that optimistically start at a lattice's bottom value may only refine 
approximations upwards; pessimistic analyses only downwards.
OPAL can automatically check the monotonicity of updates.
Monotonicity allows analyses to know which refinements of intermediate results are still possible.
Second, analyses must be \emph{scheduling independent}:
Whenever they receive the value of some other property they depend on, 
they must use the information provided by that value to compute the result of the current activation, i.e.,
they may not defer the incorporation of the newly gained information to a later 
activation of a continuation function. 
This ensures that all available information is used independent 
of whether the continuation is later scheduled for execution - 
an activation may never occur in case of cyclic dependencies.
For example, once the mutability analysis of a class \code{C} 
knows that \code{C}'s instance field \code{f} is mutable, 
it may no longer report that \code{C} could be immutable.
The developer of some analysis \code{A} must ensure that \code{A} is 
scheduling independent.

\begin{figure}
\begin{lstlisting}[numbers=left, numberstyle=\scriptsize\color{gray}\ttfamily,xleftmargin=2em, caption={Class Mutability Analysis}, label={lst:example_analysis}]
def analyze(Type theClass) = {
	[...]
	Blackboard.get(field, FieldMutability) match {
	case _: MutableField => return Result(theClass, MutableClass)
	case dependee: ImmutableField =>	
		if (!dependee.isFinal) dependees += (field -> dependee)
	}
	[...]
	Result(theClass, ImmutableClass, dependees, continuation)
}
\end{lstlisting}
\end{figure}

\subsection{Declarative Specifications}
\label{subsec:lifecycle}
On top of the IAF and CF, the developer of an analysis \code{A}
specifies (a) the property kinds computed by \code{A}, (b) its dependencies, 
(c) on which entities \code{A} will be executed 
and (d) when the blackboard should start \code{A's} 
execution. 
These specifications are evaluated when the analysis is registered with the 
blackboard, before the latter takes over control of analysis activation.
When registering analyses, developers may also report precomputed values to the blackboard (\req{external}).

The specification of the computed property kinds also states whether intermediate results
are optimistic or pessimistic and whether the analyses contributes to a collaborative computation or intends to be the only analysis computing the specified property kinds.
Dependency specifications state other property kinds on which 
\code{A} depends (which \code{A} queries) and 
whether \code{A} can process optimistic/pessimistic intermediate values
or final values only.

Analyses can eagerly select a set of entities (e.g., all methods of the analyzed program) 
if it is likely necessary to perform the analysis for all of these entities (\req{eager}).
This is, e.g., useful for analyses that are of interest to the end user, 
e.g., if the user is interested in the purity of all methods. 
Alternatively, analyses can be registered to be invoked lazily~\cite{jensen2010interprocedural,bodden2018secret}.
Lazy analyses only compute a property if that property is queried (\req{lazy}) by another analysis or by the end user.
Finally, an analysis can specify a property kind $k$ such that it is started for every entity for which $k$ has been computed (\req{triggered}).

Some analyses benefit from enforcing a specific order for computing the properties for different entities (\req{incremental}).
For instance, the class mutability analysis benefits from traversing the class hierarchy downwards, 
such that results for a parent class are available before any subclass is analyzed.
In OPAL, this is supported by enabling the developer of an analysis \code{A}
to declare a number of computations to be scheduled whenever \code{A}
returns a result to the blackboard.

\begin{figure}
\begin{lstlisting}[numbers=left, numberstyle=\scriptsize\color{gray}\ttfamily,xleftmargin=2em, caption={Registration of Class Mutability Analysis}, label={lst:registration}]
override def derivesLazily = Optimistic(ClassMutability)
override def uses = Set(Optimistic(FieldMutability))
override def register() = {
	Blackboard.set(Type.Object, ImmutableClass)
	val analysis = new ClassMutabilityAnalysis
	Blackboard.registerLazyAnalysis(this, analysis.analyze)
}
\end{lstlisting}
\end{figure}

For illustration, Listing~\ref{lst:registration} shows the registration code for a class mutability analysis.
Line \tnum{1} declares that the analysis optimistically and lazily derives class mutability.
Line \tnum{2} declares that in performing its computation, it may require field mutability and that it can handle intermediate results for this property if they were computed optimistically.
This declaration is complete: No property kinds other than field mutability (and class mutability) may be queried by this analysis.
Line \tnum{4} registers a predefined value stating that the base class \texttt{Object} is immutable (\req{external}).
The IAF \texttt{analyze} is registered as a lazy analysis in line \tnum{6}, i.e., the mutability of a certain class will only be computed 
on demand, e.g., when a purity analysis queries it.

\subsection{Reporting Results}
\label{subsec:results}
As already mentioned, analyses write intermediate and final results to the blackboard.
They can report results for each single entity individually or for multiple entities at the same time.
A result consists of a single lattice value representing the new value for the property
or of an update function (UF) for updating the property's current value 
(as recorded in the blackboard) to incorporate the new result.

A UF is used for properties whose computation is not 
localized to a specific part of the program, e.g., the callers of a method.
For such properties, constraint-based analyses~\cite{aiken1999introduction,nielson2005principles} 
have been used in the past; declarative analyses also provide such updates, called deltas, 
that only specify the change to the property value instead of the full new property value. 
The \code{UF} merges the results of one activation 
to the current state of the property (e.g., add a new caller to an existing set of callers).
This way, activations of one or of different analyses
can collaboratively contribute to a property (\req{partial}, \req{partialcollab}).

\subsection{Execution Constraints}
\label{subsec:execution_constraints}

Once the end user chooses a set of analyses to be executed (\req{pluggability}),
OPAL uses the declarative specifications (Section~\ref{subsec:lifecycle}) to 
check and automatically enforce restrictions on analyses that can be executed together.
First, it ensures that any property kind is computed by at most one analysis 
or collaboratively; this is to avoid that conflicting results are reported to the blackboard.
Second, if several analyses derive a property kind collaboratively, OPAL ensures that
they are all either optimistic or pessimistic.
Finally, OPAL ensures that all property kinds required by any analysis
are derived by another analysis or there is a fallback value provided; this is to ensure that 
dependencies can be satisfied.

OPAL's blackboard may run optimistic and pessimistic analyses simultaneously.
But, when doing so, it ensures that no intermediate results are propagated 
between them (\req{optivspessi}). 
Given property kind $p$ that is computed optimistically and pessimistic analysis $A$ depending on $p$, OPAL does not forward any intermediate values of $p$ to $A$'s CF. The latter is triggered
only when a value of $p$ is submitted marked as final. We say that the dependency of $A$ on $p$
 is \emph{suppressed}.
There are subtle interactions between dependency 
suppression and cyclic and collaborative computations, 
which we explain next.

First, there can be no cyclic dependencies between pessimistic and optimistic analyses.
The correctness of cyclic dependency resolution relies on the assumption that  
all intermediate approximations have been processed and no further 
updates to any property involved in the cycle may happen (cf.\ Section~\ref{subsec:fixed-point}). 
This obviously is not the case when updates are suppressed. 

The interaction between dependency suppression and collaboratively computed properties 
is more involved. Assume a collaboratively computed property $p_1$  that 
(transitively) depends on another collaboratively computed property $p_2$ 
and consider the case when one or more of the transitive dependencies between 
them is suppressed\footnote{On a chain of dependencies, more than one may be suppressed. Also, if $p_1$ depends on $p_3$ and $p_4$ and each of those depends on $p_2$, there is more than one path between $p_1$ and $p_2$, on which dependencies may get suppressed.}. 
In this case, OPAL must ensure that $p_2$'s values
 are committed as final before $p_1$'s values 
 can be committed as final, too. This ensures that final values have been 
propagated along the suppressed dependencies.
To this end, OPAL derives a \emph{commit order} when checking the execution constraints before executing analyses.
The commit order is a partial order between collaboratively computed property kinds:
$p_1$ must be finalized later than any other collaboratively computed property kind $p_2$
on which $p_1$ depends when there is suppression between them.

Suppression of intermediate updates can also be used to improve performance:
Consider the relation between TAC and AI.
Both are optimistic and TAC could use intermediate AI results.
But these results are typically not useful, hence, it can be beneficial to use suppression
to avoid costly computation of these intermediate results and instead compute the TAC only once on the final AI result.

\subsection{Fixed-Point Computation}
\label{subsec:fixed-point}

Computation is started for the entities selected by eager analyses (\req{eager}) (cf.\ Section~\ref{subsec:lifecycle}).
Whenever intermediate values for properties are submitted, the blackboard schedules activations of continuation functions,
distributing updated results to analyses that depend on them.
Additionally, the blackboard starts new computations by invoking the initial analysis function for properties that are requested lazily (\req{lazy}), 
are triggered by some analyses reaching a certain entity (\req{triggered}), or whenever it is guided to do so 
by running analyses (\req{incremental}).
This process of scheduling IAF and CF activations is performed until no further updates are generated -- the blackboard has reached a \emph{quiescent} state.
At this point, however, the properties' values may not necessarily be final, as there still may be unresolved dependencies.
There are three cases to be considered.

First, an analysis was scheduled for some property kind $p$, 
but it did not analyze some entity $e$, 
for which $p$ was requested, e.g., because $e$ was not reachable in the call graph. 
In this case, the \emph{default value} (\req{defaults}) is inserted, 
which may trigger further computations, until quiescence is reached again. 

Second, properties that cyclically depend on each other are not finalized yet.
If such properties form a \emph{closed strongly connected component}, i.e., they do not have any dependees outside of the cycle (but other properties may still depend on them), they are now finalized to their current value.
By requiring analyses to report their results in a monotonous and scheduling independent way 
(cf.\ Section ~\ref{subsec:analysis_structure}), OPAL guarantees that the cycle 
resolution is deterministic and sound. Again, further computations may arise from resolving cyclic dependencies 
(including supplying more default values and resolving further cycles) until quiescence is reached again.

Finally, the blackboard finalizes values for collaboratively computed properties.
It respects the \emph{commit order} computed previously (cf.\ Section~\ref{subsec:execution_constraints}):
After finalizing a set of collaboratively computed properties, computation is resumed again.
Only once quiescence is reached again, the next property kinds, as given by the commit order, are finalized.
This is repeated until all collaboratively computed properties have been finalized.

\subsection{Scheduling and Parallelization}
Blackboard systems require a control component that, upon updates of the blackboard, decides 
which knowledge sources to activate next. 
In our case, this control component determines the order in which 
activations of dependent analyses are executed and is called \emph{scheduler}.
The order in which dependent analyses are activated can 
have significant effects on performance~\cite{rodriguez2011actor}.

OPAL allows for the scheduler to be easily 
exchanged in order to select the best performing one for any chosen set of analyses.
Apart from general strategies such as first-in-first-out, 
more specific algorithms may use the dependency structure or 
the values of intermediate approximations to decide the scheduling order.
This is similar to the control component of blackboard systems 
asking knowledge sources for an estimated information gain (cf.~\cite{corkill1991blackboard}).

\label{subsec:approach_parallelization}
Blackboard systems lend themselves well to parallelization.
The individual knowledge sources, i.e., analyses in our case, are decoupled and their activations (both the initial analysis and the continuations) can be executed in parallel on multiple threads.
Updates to the blackboard, on the other hand, can be synchronized on a special thread or, 
if that becomes a bottleneck, distributed to several threads based on the property kind and/or entity.
A simple implementation may consist of several threads that use a shared data structure holding the property data and use locks or other mechanisms to synchronize accesses to this shared storage.

\subsection{Summary}
Our approach fosters strong decoupling of reified lattices (choice of data structures), analyses (choice of algorithm), and the solver infrastructure (the concrete fixed-point solving implementation).
This enables exchanging and optimizing these parts independently. 
As reified lattices are the basis for all communication between analyses, different versions of analyses can be implemented at different trade-offs.
The solver manages execution of analyses, tracks dependencies and propagates updates, performs monotonicity checks, and computes the fixed-point solution.

\section{Evaluation}
\label{section:evaluation}
We evaluate our approach by answering the following questions:
\begin{itemize}[labelindent=0.5cm,labelsep=0.1cm,leftmargin=*,noitemsep,nolistsep]
	\item[\rqd{applicability}] Does OPAL support modularization of a broad range of static analysis kinds with varying requirements?
	\item[\rqd{exchangability}] Does exchangeability of analysis modules benefit the end user and the developer?
	\item[\rqd{parallelization}] Can the framework be parallelized?
	\item[\rqd{datastructures}] What is the benefit of analysis-specific data structures?
	\item[\rqd{doopcomparison}] How does the performance of OPAL's analyses compare to state-of-the-art declarative approaches?
\end{itemize}

We implemented our approach on top of the Scala-based OPAL framework for JVM bytecode analysis~\cite{eichberg2014software}.
However, the approach is framework and language independent.
We answer the above research questions using the case studies of Section~\ref{section:case_study} to analyze the DaCapo 2006 benchmark~\cite{blackburn2006dacapo}.
We choose DaCapo because Doop, which we compare to in Section~\ref{subsec:doop_comparison}, has special support for it.
Both the implementation of OPAL as well as the case studies are available in the OPAL GitHub repository\footnote{https://github.com/stg-tud/opal}.

All measurements were performed in a Docker container\footnote{https://doi.org/10.5281/zenodo.3872848} on a server with two AMD(R) EPYC(R) 7542 @ 2.90\,GHz (32 cores / 64 threads each) CPUs and 512~GB RAM.
Analyses were run using OpenJDK 11.0.5+10 (64-bit) with 32\,GB of heap memory and Scala 2.12.9.
Experiments were run \tnum{seven} times and we report their median runtime.
We report only excerpts of the results here\footnote{The entire results can be found here: https://doi.org/10.5281/zenodo.3972736}.

\subsection{Support for Various Analyses}
To answer \rq{applicability}, we implemented the case studies from Section~\ref{section:case_study} using OPAL and argue that these are representatives of different analysis kinds. 
The first case study represents pessimistic analyses in the context of improving precision of a three-address code representation (TAC)---it shows how basic analyses can be extended by analyses that are specialized to increase the precision of sub-problems' solutions.
The modular call graph of the second case study involves tightly interacting yet decoupled analyses (e.g., points-to and call graph) and demonstrates how one can plug in further modular analyses that handle special cases of Java in order to increase the call graph's soundness.
The third case study introduced several exchangeable analyses for different high-level properties (immutability, escape information, purity).
The individual analyses are relatively simple and can focus on their respective property, but by using the results of other analyses,
they can be more precise than a corresponding monolithic analysis of medium complexity.

As discussed in Section~\ref{section:case_study}, to achieve this modularity, several requirements need to be satisfied
(cf. Table~\ref{tbl:req}).
Section~\ref{section:approach} already explained how OPAL supports all of them.
On the contrary, as we argue in Section~\ref{sec:req}, no current imperative or declarative framework supports all these requirements.

We additionally
implemented a solver for \emph{inter-procedural, finite, distributive subset problems} (IFDS)~\cite{reps1995precise}, a well-known general framework for dataflow problems based on graph reachability.
Similar to other IFDS solvers, e.g., Heros~\cite{Heros}, users provide a domain for their dataflow facts and four flow-functions that together specify the IFDS problem.
The solver starts one computation per pair of method and entry dataflow fact and these tasks need to communicate their results.
We chose IFDS as it is a general framework that allows implementing many dataflow analyses and it is dissimilar from the \tnum{three} case studies' analyses.
In particular, it shows OPAL's support for implementing general solvers as individual analyses.

\observation{OPAL's programming model enables the implementation of dissimilar analyses, fostering their modularization into a set of comprehensible, maintainable, and pluggable units. OPAL is the only static analysis framework satisfying all requirements from Section~\ref{sec:req}.}

\subsection{Effects of Exchangeability of Analyses}
Our approach 
strictly decouples property kinds from analyses computing them.
Thus, it can provide different analyses computing the same property kind to cover a wide range of precision, sound(i)ness, and performance trade-offs.
Two experiments examine how this exchangeability fosters 
rapid probing, thus benefiting the analysis' developer and end user alike (\rq{exchangability}):
We explore the impact on precision in one experiment and that on soundness in the second. 

\begin{table}
	\footnotesize
	\centering
	\caption{Purity results for different configurations (hsqldb)}
	\begin{tabularx}{\columnwidth}{Xrrrrr}
		\toprule
		\textbf{Configuration} & \textbf{\#Pure} & \textbf{\#SEF} & \textbf{\#Other} & \textbf{\#Impure} & \textbf{\faClockO{} Analysis}\\
		\midrule
		$\text{PA}_2$/$\text{FMA}_1$/$\text{E}_1$ 		&417	&482	&245	&\numprint{2635}	&2.42\,s\\
		$\text{PA}_2$/$\text{E}_1$		 				&363	&536	&245	&\numprint{2635}	&2.40\,s\\
		$\text{PA}_2$/$\text{FMA}_1$/$\text{E}_0$ 		&417	&481	&241	&\numprint{2640}	&1.93\,s\\
		$\text{PA}_2$									&362	&504	&225	&\numprint{2688}	&0.98\,s\\
		$\text{PA}_1$/$\text{FMA}_1$					&415	&431	&0		&\numprint{2933}	&0.93\,s\\
		$\text{PA}_0$/$\text{FMA}_1$					&104	&0		&0		&\numprint{3675}	&0.70\,s\\
		$\text{PA}_0$									&100	&0		&0		&\numprint{3679}	&0.13\,s\\
		\bottomrule
	\end{tabularx}
	\label{tbl:exch_purity}
\end{table}

In our first experiment, we run 
various configurations of our purity analysis (cf.\ Section~\ref{sec:purity}) with different supporting analyses for field mutability or escape information with different precision-scalability trade-offs.
No other tool supports similar exchangeability of interacting purity, mutability, and escape analyses.
Table~\ref{tbl:exch_purity} shows the results for \emph{hsqldb}. 
Higher indices indicate more precise analyses.
Comparing the least precise analysis $\text{PA}_0$ with the most precise $\text{PA}_2$/$\text{FMA}_1$/$\text{E}_1$, we observe a reduction in the number of reported impure methods by \textasciitilde\tnum{28}\%, but a runtime slowdown by \tnum{18.6}x. Some configurations even have a large impact on runtime for almost no gain in precision, e.g., comparing the most precise one with that using simpler escape analysis E$_0$.

\begin{table}
	\footnotesize
	\centering
	\caption{Results for different call graph modules for Xalan}
	\begin{tabularx}{\columnwidth}{Xrrrr}
		\toprule
		\textbf{Configuration} 		& \textbf{\#Reachable Methods} & \textbf{\#Edges} & \textbf{\faClockO{} Analysis}\\
		\midrule
		\text{RTA}					&\numprint{6141}		&\numprint{46946}		&8.58\,s\\
		\text{RTA\_C}				&\numprint{6162}		&\numprint{47154}		&8.76\,s\\
		\text{RTA\_R}				&\numprint{8404}		&\numprint{63821}		&10.07\,s\\
		\text{RTA\_X}				&\numprint{12937}		&\numprint{106516}		&12.99\,s\\
		\text{RTA\_C\_X}			&\numprint{12958}		&\numprint{106743}		&12.86\,s\\
		\text{RTA\_S\_T\_F\_C\_X}	&\numprint{12970}		&\numprint{106778}		&13.35\,s\\
		\bottomrule
	\end{tabularx}
	{C=Configured native methods; R=Reflection; X=Tamiflex;\\ S=Serialization; T=Threads; F=Finalizer;}
	\label{tbl:exch_cg}
\end{table}

In the second experiment, we evaluate the RTA call graph with different supporting modules for different JVM features.
While DOOP computes call graphs and offers some modularity, e.g., for reflection, no other tool so far includes such fine-grained modules for call graphs.
Also, DOOP does not support RTA, but points-to based call graphs only.
Results for \emph{Xalan} are shown in Table~\ref{tbl:exch_cg}, displaying the active modules, the number of reachable methods (RMs), 
call edges, and respective construction time.
While some configurations discover more methods/edges than others, they may discover different sets of methods/edges.
A configuration is only guaranteed to be strictly more sound if it uses a strict superset of modules.
Compared to the baseline, \text{RTA} with support for preconfigured native methods (\text{RTA\_C}), reaches \tnum{21} more methods and \textasciitilde\tnum{200} more call edges. 
Reflection support (\text{RTA\_R}) brings over \tnum{\numprint{2000}} more RMs and \tnum{\numprint{16000}} call edges; at the same time, construction time increases by about \tnum{15\%}.
Using the Tamiflex (\text{RTA\_X}) module instead
increases call graph size (and soundness) more but introduces further slowdown.
With all modules enabled, we reach \tnum{111\%} more methods and \tnum{127\%} more call edges, 
at the cost of a \tnum{55}\% increased runtime.
Moreover, the data
suggests that different modules benefit different projects.
Tamiflex impacted \emph{Xalan} and \emph{jython}, reflection \emph{fop},
and serialization \emph{hsqldb}. 
Thus, which modules are more relevant than others may differ between different programs
and it may be worth investigating tradeoffs even at the level of individual projects. 

Overall, both experiments confirm that OPAL maintains exchangeability benefits 
from Datalog-based analyses, while generalizing these results to a broader range of lattices.

\observation{OPAL facilitates systematic investigation of different configurations, supporting users and developers in finding the best trade-off between precision, sound(i)ness, and scalability.}

\subsection{Parallelization}
\label{sec:parallel}

\begin{figure}
	\centering
	\includegraphics[width=\columnwidth]{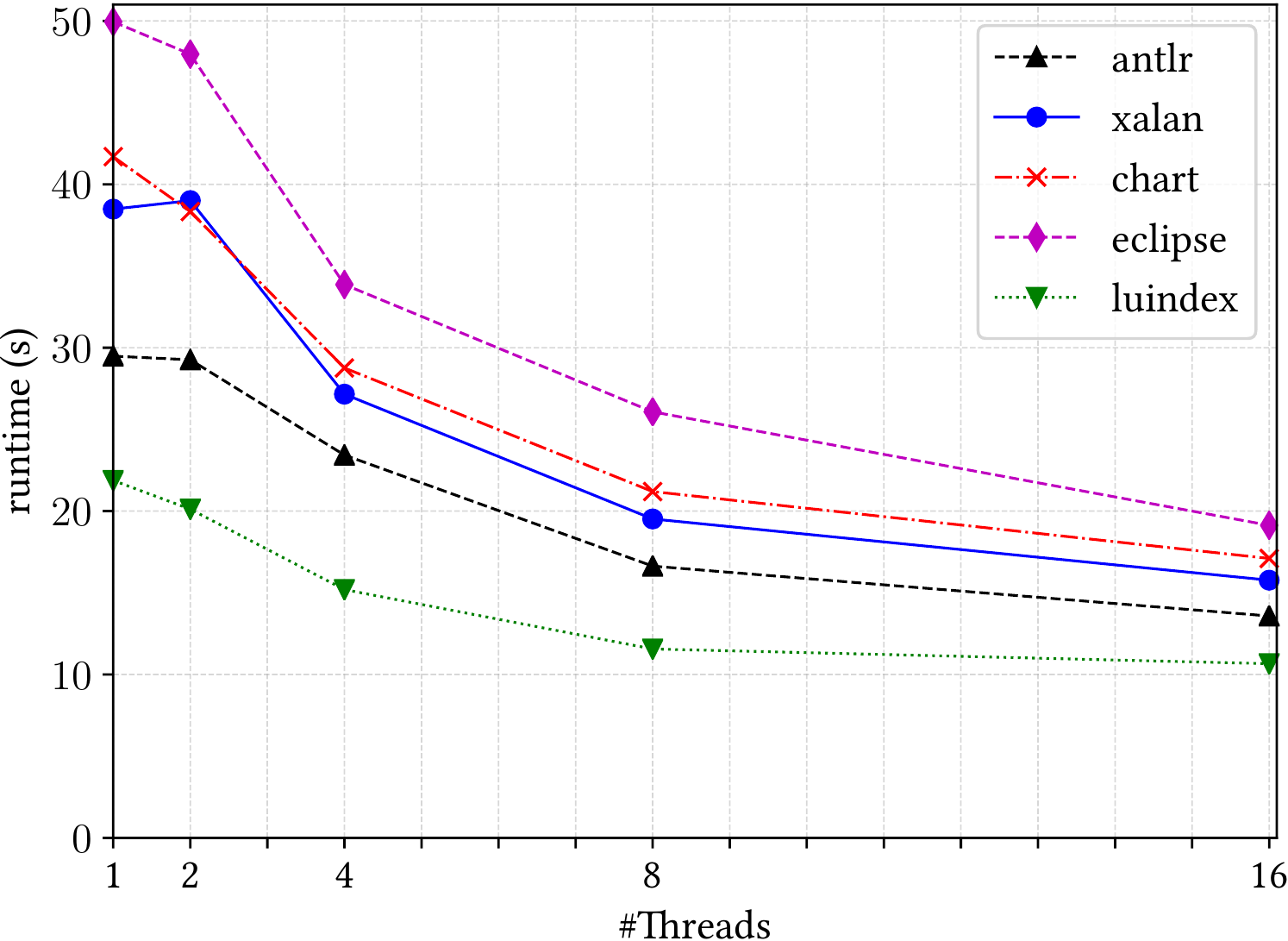}
	\caption{Parallel architecture performance}
	\label{fig:parallelization}
\end{figure}

We implemented a proof-of-concept parallel version of our blackboard control (\rq{parallelization}).
Using this, we measured the execution time for the points-to-based call graph with different numbers of threads.
Results for five DaCapo projects are shown in Figure~\ref{fig:parallelization}.
The projects were selected to have similar runtime to facilitate graph readability, the other projects show similar behavior.
Benefits of parallelization over one thread appear at two to four threads and we achieve speedups of up to \tnum{2}x for \tnum{16} threads.
Beyond this, further improvement is negligible; instead, it slightly decreases due to growing communication overhead.
These results are encouraging, given that the parallel version is not at all optimized. 
An optimized version of it is expected to scale better.
Designing such an optimized version requires further research to identify the optimal way to parallelize the computation.

\observation{OPAL's computation can be parallelized and that parallelization holds potential for increased performance.}

\subsection{Benefits of Specialized Data Structures}
To answer \rq{datastructures}, we compare two versions of the same points-to based call-graph algorithm.
Both encode points-to, caller, and callee information as integer values.
The first version uses specialized trie-based data structures,
the second one uses standard Scala sets.

Results are given in the sixth and last column of Table~\ref{tbl:doop_comparison}.
Due to its high memory consumption, we had to run the version using Scala's data structures with 128~GB of heap space; \emph{jython}'s analysis even required 256~GB.
Using tailored data structures, OPAL's runtime decreased by \tnum{65\%} to \tnum{98\%} compared to naively using Scala's sets.

\observation{Selecting suitable data structures adapted to the specific analysis needs is an important factor for analysis performance. While the analysis developer can freely select optimized data structures in OPAL, strictly declarative approaches do not support such choices.}  

\begin{table}
	\footnotesize
	\caption{Runtime and size of points-to based call graphs}
	\setlength{\tabcolsep}{4.5pt}
	\begin{tabularx}{\columnwidth}{Xrrrrrrr}
		\toprule
		& \multicolumn{4}{c}{\textbf{DOOP}}	& \multicolumn{2}{c}{\textbf{OPAL}}	& \multicolumn{1}{c}{~\textbf{OPAL}}\\
		\cmidrule(lr){2-5} \cmidrule(lr){6-7}
		\textbf{Project}	& Compile & Facts & Analysis & \#RM		&	runtime & \#RM			&\multicolumn{1}{c}{~(Scala)}\\
		\midrule
		antlr 				& 107\,s		& 35\,s		& 41\,s		& \numprint{8402}	& 28.36\,s	& \numprint{8653}	& 305.90\,s\\
		bloat 				& 109\,s		& 21\,s		& 33\,s		& \numprint{9644}	& 34.43\,s	& \numprint{10000}	& 266.08\,s\\
		chart 				& 109\,s		& 38\,s		& 45\,s		& \numprint{12058}	& 40.13\,s	& \numprint{12268}	& 516.37\,s\\
		eclipse 			& 109\,s		& 19\,s		& 17\,s		& \numprint{7163}	& 44.89\,s	& \numprint{13429}	& 343.69\,s\\
		fop 				& 110\,s		& 41\,s		& 35\,s		& \numprint{7300}	& 18.87\,s	& \numprint{7509}	&  56.64\,s\\
		hsqldb 				& 109\,s		& 38\,s		& 32\,s		& \numprint{7097}	& 19.65\,s	& \numprint{7455}	&  55.69\,s\\
		jython 				& 108\,s		& 24\,s		& 90\,s		& \numprint{12901}	& 77.65\,s	& \numprint{13161}	& \numprint{3341.62}\,s\\
		luindex 			& 108\,s		& 21\,s		& 19\,s		& \numprint{7608}	& 19.34\,s	& \numprint{7972}	&  62.57\,s\\
		lusearch 			& 108\,s		& 21\,s		& 20\,s		& \numprint{8281}	& 21.03\,s	& \numprint{8540}	&  70.55\,s\\
		pmd					& 109\,s		& 39\,s		& 36\,s		& \numprint{8817}	& 21.47\,s	& \numprint{9028}	&  75.47\,s\\
		xalan				& 108\,s		& 37\,s		& 30\,s		& \numprint{7111}	& 35.59\,s	& \numprint{13330}	& 246.97\,s\\\midrule
		geo. $\varnothing$	& 108.54\,s	& 29.09\,s	& 32.51\,s	&	& 29.68\,s	&	& 191.26\,s\\
		\bottomrule
	\end{tabularx}
	\label{tbl:doop_comparison}
\end{table}

\subsection{Comparison with Declarative Approaches}
\label{subsec:doop_comparison}
After evaluating individual unique features of OPAL in isolation, we present the results of an experiment that directly compares the performance of OPAL with that of Doop~\cite{bravenboer2009strictly} (\rq{doopcomparison}) -  
a highly optimized state-of-the-art 
tool for declarative Java points-to and call-graph analyses on top of
the Soufflé~\cite{jordan2016souffle} Datalog engine.
Its declarative approach assembles a fair comparison 
as it supports similar modularity and configurability and good trade-offs between pluggable precision/recall.
Also, Doop's and Soufflé's authors repeatedly claimed its good performance~\cite{bravenboer2009strictly,smaragdakis2010using,bravenboer2009exception,jordan2016souffle}.
Specifically, we compare our points-to based call-graph's runtime from Section~\ref{sec:call_graph} to Doop's.

For better comparability, we disabled the reflection support in both tools, because the respective approaches are different.
The applications were analyzed together with OpenJDK 1.7.0\_75 (used for the TamiFlex data in Doop's benchmarks).
Minor differences (less then 5\% difference in the number of RMs, except for eclipse and xalan) remain, but these are in Doop's favor, 
since they result in more work to be done by OPAL\footnote{For instance, OPAL does handle some cases of reflection more soundly even with reflection handling disabled in order to process the DaCapo benchmark correctly.}.
Still, the sixth column of Table~\ref{tbl:doop_comparison} shows that our complete analysis, including all preprocessing, is often faster than Doop's analysis (\tnum{9\%} in the geometric mean).
Further, Doop additionally requires time for rule compilation and fact generation.

We used OPAL's single-threaded implementation since it seems that 
Doop is hardly parallelized (fact generation was done with \tnum{128} threads, 
but did not significantly vary with other values for the \texttt{fact-gen-cores} 
parameter and the \texttt{souffle-jobs} parameter did not show any effects).
Using a parallel version, OPAL should 
be able to outperform Doop even more as shown in Section~\ref{sec:parallel}.

\observation{Despite being more general, i.e., not tuned for points-to analyses but supporting many different kinds of analyses, OPAL clearly outperforms Doop.}
\section{Related Work}
\label{section:related_work}

In this section, we discuss several related approaches in various areas of static analysis as well as in blackboard systems.

\subsection{Blackboard Systems}
The blackboard metaphor was introduced by Newell~\cite{newell1962some} and implemented for speech-recognition in HEARSAY-II~\cite{erman1980hearsay}.
Blackboard systems were used for image recognition~\cite{li1995object}, vessel identification~\cite{nii1982signal}, or industrial process control~\cite{dodd2009industrial}.
For these domains, no efficient, deterministic algorithm is known, leading to several problems mentioned by Buschmann et al.~\cite{buschmann1996patternoriented}:
nondeterminism making testing difficult, no guarantee for good solutions, performance suffering from wrong hypotheses, and high development effort due to ill-defined domains.
As static analyses have a well-defined domain and deterministic algorithms, these do not apply to our approach.

The structure of blackboard systems is described, e.g., by Nii~\cite{nii1986blackboard}, Craig~\cite{craig1988blackboard}, and Corkill~\cite{corkill1991blackboard}.
Corkill also discusses concurrent execution of knowledge sources and the control component~\cite{corkill1988design}, similar to OPAL.
OPAL resembles a more modern interpretation of blackboard systems~\cite{craig1993new}:
its blackboard is not hierarchical
and analyses may keep state between activations.
Information is, however, never erased and all communication is done via the blackboard.

Brogi and Ciancarini used the blackboard approach to provide concurrency for their Shared Prolog language~\cite{brogi1991concurrent}.
Like static analyses, this domain is well-defined.
Their knowledge sources are restricted to be Prolog logical programs, 
while OPAL's analyses can be implemented in a way best suited to the analysis needs.

Decker et al.~\cite{decker1991effects} discuss the importance of heuristics for scheduling concurrent knowledge source activations.
Focusing on static analyses and well-defined dependency relations, 
OPAL provides good general heuristics which are agnostic to individual analyses.

\subsection{Abstract Interpretation}
Cousot et al.~\cite{cousot1979systematic} have proven that multiple (possibly cheap) abstract domains (i.e., analyses) can be combined using the reduced product to increase overall precision.
In abstract interpreters, such as Astr\'ee~\cite{cousot2006combination} or Clousot~\cite{fahndrich2010static}, dependencies between domains are restricted by the execution order.
Thus, the same program statement must be analyzed multiple times which is superfluous with OPAL's explicit dependency management.
Also, abstract interpretation typically aims to compute abstract \emph{approximations}~\cite{cousot1977abstract} of concrete values, such as an integer variable's value.
OPAL further allows natural expression of analyses on all granularity levels.
Keidel et al.~\cite{keidel2018compositional,keidel2019sound} provide modular and reusable abstract semantics 
for different language features allowing soundness proofs from composition of already sound components.
The analyses again approximate single concrete program values.
OPAL supports analyses to be based on abstract interpretation and includes such analyses, 
but generalizes to a much broader range of static analyses.

\subsection{Declarative Analyses Using Datalog}
Datalog is often used to implement static analyses in a strictly declarative fashion~\cite{reps1995demand,whaley2004cloning,lam2005context,hajiyev2006codequest,whaley2007context,eichberg2008defining}.
Properties are represented as relations and rules specify how to compute them.
This enables modularization, as rules can be easily exchanged and/or added (e.g.\ for new language features).
The Doop~\cite{bravenboer2009strictly} framework, building on top of the highly optimized Datalog solver Souffl\'e~\cite{jordan2016souffle}, has shown that the rule-based approach enables precise and scalable points-to analyses.
For this reason, Doop became the state-of-the-art for such analyses~\cite{smaragdakis2011pick, kastrinis2013hybrid, smaragdakis2015more, tan2016making, tan2017efficient}.
Datalog-based frameworks, however, are limited in their expressiveness by using relations, i.e., set-based abstractions, to represent all analysis results.
OPAL's approach combining imperative and declarative features provides similar benefits as Datalog-based approaches, while allowing for more expressive ways to represent data and to implement analyses.

Datalog's limitation to relations has also been pointed out by Madsen et al.~\cite{madsen2016datalog}.
They propose Flix to overcome this using a language inspired by Datalog and Scala to specify declarative pluggable analyses using arbitrary lattices as in OPAL.
However, Flix focuses on verifying soundness and safety properties of static analyses and not on performance.
For instance, Flix does not allow optimized data structures or scheduling strategies.
We wanted to compare our approach against Flix and contacted the authors, but they answered that their IFDS implementation is dysfunctional now and suggested comparing against Doop with the Soufflé engine, which we did in Section~\ref{subsec:doop_comparison}.
Szabó et al.~\cite{szabo2018incrementalizing} also extend Datalog to allow arbitrary lattices for static analysis.
Their solver IncA focuses on incrementalization.
OPAL allows optimizations, e.g., of used data structures or scheduling strategies.
Furthermore, analyses' coarser granularity compared to individual rules reduces overhead in parallelization.

\subsection{Attribute Grammars}
Attribute grammars~\cite{knuth1968semantics} used in compilers such as JastAdd~\cite{ekman2007jastadd} 
enable modular inference of program properties by adding computation rules to the nodes of a program's abstract syntax tree (AST).
In traditional attribute grammars, attributes may only depend on parent, sibling, and child nodes.
Circular reference attribute grammars~\cite{farrow1986automatic, jones1990efficient, hedin2000reference, magnusson2007circular} enable attributes to depend on arbitrary AST nodes and allow circular dependencies.
Still, analyses are tightly bound to the AST, impeding natural expression of analyses based on different structures, such as a control-flow graph.
Similar to OPAL, JastAdd enables pluggability for new language features.
However, JastAdd requires at least one attribute in a cyclic dependency to be marked explicitly, while OPAL handles this transparently.

\"Oqvist and Hedin~\cite{oqvist2017concurrent} proposed concurrent evaluation of low complexity attributes in circular reference attribute grammars.
OPAL on the other hand supports arbitrary granularity of concurrent computation.
OPAL's explicit dependency management
enables analyses to drop dependencies and commit final results early for improved performance.
Finally, as memorization of properties is done in OPAL's blackboard, temporary values are garbage collected automatically,
whereas JastAdd requires explicit removal.

\subsection{Imperative Approaches and Parallelization}
Lerner et al.~\cite{lerner2002composing} proposed modularly composed dataflow analyses which communicate implicitly through optimizations of the analyzed code or explicitly through \emph{snooping}. 
A fixed-point algorithm repeatedly reanalyzes the code, while OPAL's explicit dependencies avoid reanalysis.
They support only dataflow analyses, while OPAL enables a wide range of analyses including dataflow analyses.

CPAchecker~\cite{beyer2007configurable} is a tool for configurable software verification and analysis.
For any combination of analyses, CPAchecker requires defining a compound analysis to integrate results of individual analyses and manage their interaction.
For CPA+~\cite{beyer2008program}, combined analyses must work with the same domain and provide an explicit measure of result precision.
In contrast, OPAL enables tight interaction and interleaved execution of independently-developed analyses without requiring a compound analysis or explicit measure of precision.

Johnson et al.~\cite{johnson2017collaborative} present a framework for collaborative alias analysis.
Clients ask queries which are processed by a sequence of analyses.
Each analysis can answer the query or forward it to the next one.
Analyses can also generate additional (premise) queries.
To ensure termination, a complexity metric must be defined and premises must be simpler than the queries they originate from.
Therefore, cyclic dependencies, required for optimal precision, and
results combined from different analyses are not supported.

Parallel execution of static analyses is performed by Magellan~\cite{eichberg2006integrating}.
In this framework, dependencies are given by the data processed instead of explicitly by the analyses.

Haller et al.~\cite{haller2016reactive} concurrently execute tasks based on lattices and apply this to static analysis.
Their framework requires dependencies to be managed fully by the client while OPAL manages them automatically based on declarative specifications. 
In recent work~\cite{helm2020programming}, we extended this approach to support mutable state and found that exchangeable scheduling strategies significantly impact performance.
Both concepts are supported in OPAL.
\section{Threats to Validity}
\label{section:threats_to_validity}

One threat to the validity of our evaluation is the use of the relatively old and small DaCapo benchmark.
It is, however, widely used to evaluate Doop~\cite{bravenboer2009strictly} and to compare other approaches with Doop~\cite{ali2012application,ali2013averroes,pek2014explicit,tan2016making}.
Doop's special support for the benchmark makes it a particularly fair evaluation set.
Furthermore, our experiment design, based on relative comparisons, should yield the same results with any well-assembled benchmark.

Also, our results are threatened if our points-to analysis is not sufficiently similar to Doop.
To achieve comparability, we tailored our points-to analysis to be as similar as possible, i.e., the call graph derived from the points-to results should be almost identical.
In order to ensure this, we systematically studied Doop's Datalog rules,
validated the resulting call graphs using Judge~\cite{reif2019judge} and manually inspected points-to sets from deviating call graphs.
\section{Conclusion}
\label{section:conclusion}

We presented a novel approach for modular collaborative static analyses implemented in the OPAL framework. 
Like with declarative frameworks such as Doop, 
OPAL's analyses,
while developed in isolation, can be easily composed to complex analyses by 
collaboratively computing results during interleaved executions.
Sub-analyses can be reused in various complex analyses 
and one can easily exchange sub-analyses of a complex analysis
for fine-tuning precision, sound(i)ness, and performance.

But, instead of relying on a general-purpose solver, 
OPAL combines imperative and declarative features to overcome 
limitations of fully declarative frameworks. Individual analyses can be 
implemented in imperative style making use of whatever 
data structures and implementation strategies are appropriate for their specific
needs. Interdependencies and other characteristics important 
for guiding their interleaved execution are declaratively specified and 
automatically managed by a custom solver resembling a 
blackboard architecture. Due to its approach, OPAL (a) is more general 
in terms of the analyses supported - it is in particular the first framework 
to explicitly support lazy collaboration of optimistic and pessimistic analyses - 
and (b) enables analysis-specific optimizations, which lead to
outperforming state-of-the-art declarative analyses.

We plan to explore several further research questions in the future. 
First, our evaluation suggests that better
parallelization strategies for OPAL are needed.
Second, we plan to explore further scheduling strategies, both general and analysis specific, e.g., strategies that abort computations whose results are no longer of interest,
or strategies (as well as analyses) that 
adapt their behavior during the execution to increase performance with 
minimal impact on precision and/or soundness. 
Last but not least, we will develop a formal model of OPAL and formally prove its properties. 
For instance, we believe that OPAL's design enables compositional soundness 
proofs~\cite{keidel2018compositional,keidel2019sound} - this needs to be proved in the future. 

\begin{acks}
This work was supported by the DFG as part of CRC 1119 CROSSING, by the German Federal Ministry of Education and Research (BMBF) as well as by the Hessen State Ministry for Higher Education, Research and the Arts (HMWK) within their joint support of the National Research Center for Applied Cybersecurity ATHENE.
\end{acks}

\bibliography{fpcf}


\end{document}